\documentclass[%
 reprint,
superscriptaddress,
 amsmath,amssymb,
 aps,
floatfix,longbibliography
]{revtex4-2}

\usepackage{graphicx}
\usepackage{dcolumn}
\usepackage{bm}
\usepackage{siunitx}
\usepackage{float}
\usepackage{natbib}
\usepackage{appendix}
\usepackage{ulem}
\usepackage{comment}
\usepackage[pagewise]{lineno}

\begin{document}

\preprint{APS/123-QED}
    
\title{Density-dependent flow generation in active cytoskeletal fluids}

\author{Tomoka Kashiwabara}
\affiliation{Department of Chemical Engineering, Kyoto University, Nishikyo-ku, Kyoto 615-8510, Japan}

\author{Tatsuya Fukuyama}
\affiliation{Department of Chemical Engineering, Kyoto University, Nishikyo-ku, Kyoto 615-8510, Japan}

\author{Yusuke T. Maeda}\email{maeda@cheme.kyoto-u.ac.jp}\affiliation{Department of Chemical Engineering, Kyoto University, Nishikyo-ku, Kyoto 615-8510, Japan}

\date{\today}

\begin{abstract}
The actomyosin cytoskeleton, a protein assembly comprising actin fibers and the myosin molecular motor, drives various cellular dynamics through contractile force generation at high densities. However, the relationship between the density dependence of the actomyosin cytoskeleton and force-controlled ordered structure remains poorly understood. In this study, we measured contraction-driven flow generation by varying the concentration of cell extracts containing the actomyosin cytoskeleton and associated nucleation factors. We observed continuous actin flow toward the center at a critical actomyosin density in cell-sized droplets. Notably, the actin flow demonstrated emergent oscillations, in which tracer advection periodically changed in a stop-and-go fashion in bulk solution. Near the onset of flow, the viscous drag of the actin cytoskeleton decreased the particle transport, after which it rapidly increased. These results indicate that the density-dependent actin flow is generated by a balance between viscous friction and contractile force.
\end{abstract}

\maketitle

\section{Introduction}

The actin cytoskeletal fibers and myosin molecular motors generate force in living cells \cite{howard1997,pollard2000, murrell2015, lee2021}. Actin fibers, which are bundled and entangled with cross-linking proteins, confer viscoelastic properties in the intracellular space and near the plasma membrane \cite{zakharov2024}. Myosin motor proteins bind to the actin network to generate contractile force through ATP hydrolysis \cite{lappalainen2022biochemical}. These proteins form the actomyosin cytoskeleton, which is responsible for processes such as cytokinesis \cite{sedzinski2011}, cytoplasmic fluid streaming \cite{brangwynne2008,ueda2010myosin}, cell deformation \cite{koenderink2018}, cell motility \cite{gardel2010mechanical, paluch2016}, and organelle positioning \cite{almonacid2015}. Understanding the formation of ordered structures within this spontaneously contracting actomyosin cytoskeleton elucidates non-equilibrium mechanics in biological complexes at the cellular level \cite{bois2011, prost2015} and provides insights into the development of force-generating polymer gels in biomolecular systems \cite{berret2016,jia20223d,saito2017understanding}.

Although advances have been made in identifying proteins involved in the actomyosin cytoskeleton and their biochemical regulation, quantitative analysis of ordered pattern formation and contraction dynamics remains challenging. Because complex regulatory networks by protein signaling and the presence of organelles make the spatial homogeneity of living cells disturbed, physical analysis of simplied cytoskeletal system is one of central issue in physics and chemistry of actomyosin mechanics. This issue is addressed through a reconstituted system where the actomyosin cytoskeleton and relevant proteins are extracted and confined to cell-sized droplet compartments. These lipid membrane-covered droplets control a dense actin network and the aster-like pattern of the microtubules, which mimic the spatial distribution of cytoskeletal structures in the intracellular space \cite{pinot2009effects, pinot2012,shah2014,schuppler2016,tan2018,suzuki2017spatial}, thus termed artificial cells. Within these artificial cells, actin fibers form a ring-like structure similar to that of the myosin contractile ring during cell division \cite{miyazaki2015,litschel2021reconstitution}. Actin polymerization facilitates flow from the membrane to the compartment \cite{malikgarbi2019,ierushalmi2020,sakamoto2020tug}. In addition to the increased contractility of the myosin motor, the polymerization rate of actin fibers is a critical parameter in these reconstitution dynamics \cite{sakamoto2023state,krishna2024}. Moreover, as the interaction between actin fibers and the lipid membrane becomes stronger, the resulting frictional force from actomyosin contraction facilitates the self-propelled migration of artificial cells \cite{sakamoto2022geometric}. These studies provide valuable insights into the density of cytoskeletal and motor proteins abundant in the cytosolic space, which is essential for generating such dynamics.

High intracellular protein concentrations maintain cellular activity, and their metabolic activity loss leads to increased viscosity \cite{guo2014}. Therefore, exploring the relationship between protein concentration and intracellular structure formation is vital for understanding the mechanisms that sustain a highly active environment. However, previous analyses of these artificial cells at near-physiological concentrations of actin cytoskeleton and myosin proteins have not fully explained the mechanisms by which emergent actin flows occur in a density-dependent manner.

\begin{figure*}[htb]
\begin{center}
\includegraphics[width=15cm]{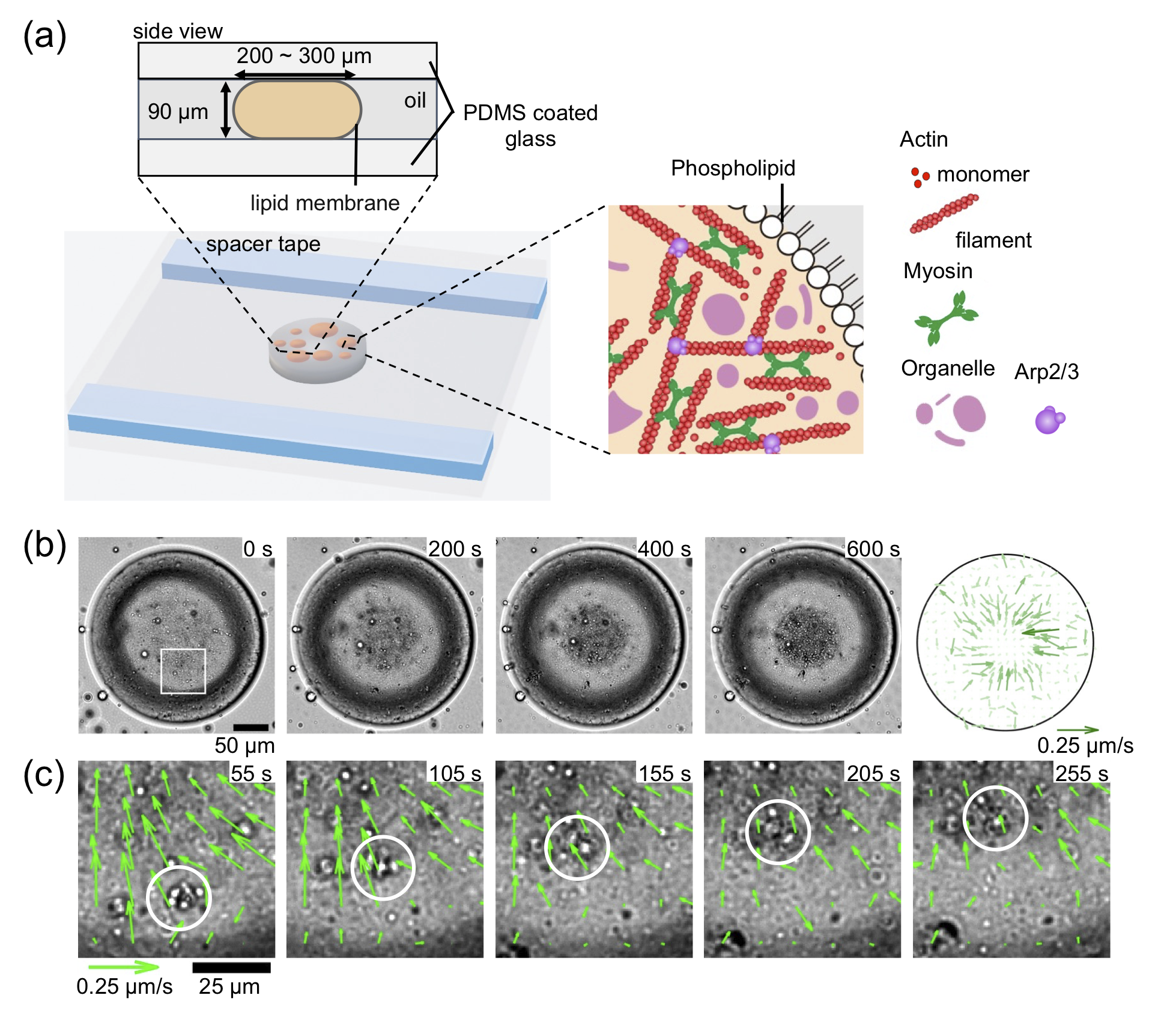}
\caption{{\footnotesize \textbf{Flow generation inside an artificial cell with actomyosin cytoskeleton}. (a) Schematic of the experimental setup. (b) Microscopic time-lapse observation of artificial cells containing actomyosin. The number in the upper right corner of each image is the elapsed time. Scale bar: \SI{50}{\micro\meter}. The figure on the right depicts the flow field of the transport of granular particles obtained using particle image velocimetry (PIV) analysis. Scale arrow is \SI{0.25}{\micro\meter\per\second}.(c) Magnified view of the transport of the granular particle toward the center, indicating that the transport direction is consistent with the flow field obtained using PIV. Scale bar: \SI{25}{\micro\meter}.}}\label{fig1}
\end{center}\end{figure*}

In this study, we experimentally examined the density dependence of actin flow generation in a reconstituted actomyosin cytoskeleton. We investigated the transition from Brownian motion to actin flow when actin density reached a critical value, as well as the effects of cell-size compartmentalization and chemical perturbation of myosin contractile activity on flow generation. This density-dependent flow of actin has implications for the higher molecular density required for cellular activity to perform functions from cytoplasmic streaming to cell migration.

\section{Results}

\begin{figure*}[tb]
\begin{center}
\includegraphics[width=17cm]{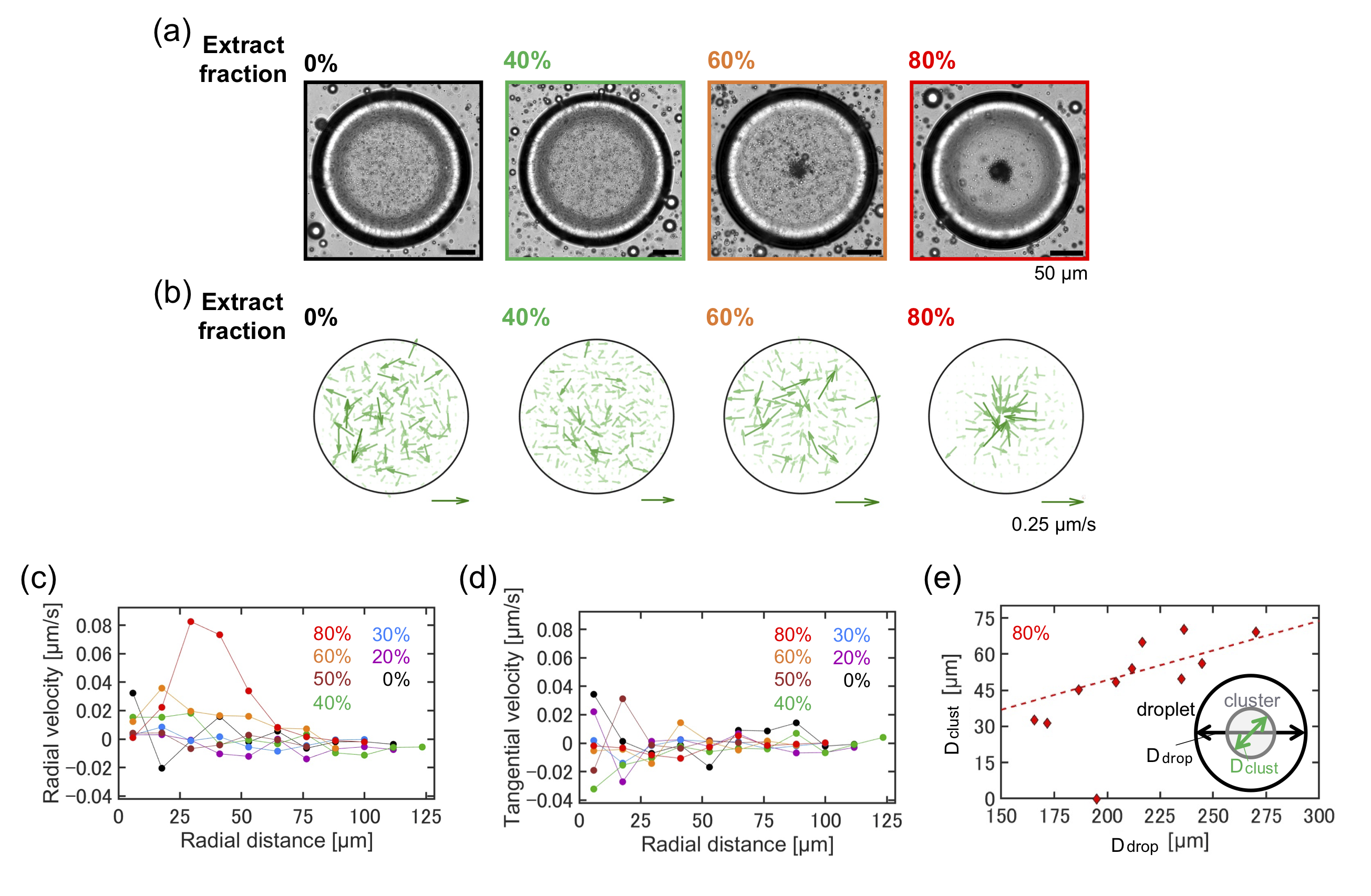}
\caption{{\footnotesize \textbf{Density dependence of actin flux inside the artificial cell.} (a) Cluster formation in artificial cells at various actomyosin density fractions ($\phi=0$\%, 40\%, 60\%, and 80\%). Scale bar: \SI{50}{\micro\meter}. (b) Actin flow fields across different actomyosin density fractions. Scale arrow: \SI{0.25}{\micro\meter\per\second}. (c) Radial velocity profile of actin flow, with radial distances measured every \SI{11.8}{\micro\meter}, which corresponds to the grid size used in the PIV analysis. (d) Tangential velocity profile of actin flow. (e) Correlation between the size of the artificial cell (extract-in-oil droplet) and the cluster size. The dashed line in red ($D_{clust} = 0.25  D_{drop}$) is drawn by least square fitting.}}\label{fig2}
\end{center}
\end{figure*}

We extracted a cytoplasmic solution containing nearly physiological concentrations of actin, myosin, and other associated proteins such as Arp2/3 from \textit{Xenopus} frog eggs \cite{sakamoto2020tug,sakamoto2022geometric,sakamoto2023state}. When myosin binds to the actin filamentous network, the actomyosin cytoskeleton is formed, which spontaneously generates contractile force coupled with ATP hydrolysis (Fig. \ref{fig1}(a)). Additionally, Arp2/3 is activated by WASP family proteins at the membrane interface and then lead to the nucleation of branched actin fibers. By enclosing this extract at various densities inside emulsified extract-in-oil droplets, the actomyosin cytoskeleton is confined within the cell-sized space at the interface of the phospholipid monolayer (Fig. \ref{fig1}(a)). We refer to this actomyosin-containing droplet as an artificial cell. A chamber for microscopic observation was constructed between two polydimethylsiloxane (PDMS)-coated glass slides, held at a height of \SI{90}{\micro\meter} with a spacer tape and artificial cells of 150-\SI{300}{\micro\meter} diameter were observed with an inverted microscope (see Methods). The artificial cells are cylindrical with a height:diameter aspect ratio of 0.3–0.6; the geometric constraints are consistent along the height, and the artificial cells can be considered as circular shapes in the horizontal plane. The constructed artificial cells also contained small granular particles, which are fragments of organelles from the extract solution. These particles served as tracers to observe actomyosin cytoskeleton dynamics using time-lapse measurements (Fig. \ref{fig1}(b)). We observed granular particles flowing from the membrane boundary toward the center of the artificial cell, forming a spherical cluster at the center (Supplementary Movie 1).

To quantitatively analyze the motion of the granular particles, we performed PIV analysis to visualize the velocity field of the flow. PIV analysis revealed that the flow toward the center of the artificial cell was dominant (Fig. \ref{fig1}(b), right). The fastest flow was observed near the cluster at the center of the artificial cell. The velocity field of the flow, together with the magnified image, indicates that the granular particles moved from near the membrane boundary with a typical velocity of approximately \SI{0.07}{\micro\meter\per\second} at approximately \SI{50}{\micro\meter} from the center of the artificial cell (Fig. \ref{fig1}(c)).

To confirm that the observed flow toward the center originated from the actomyosin cytoskeleton, we analyzed changes in the flow of particles by varying the fraction of actomyosin enclosed in the droplet by dilution. For flow analysis of the diluted extracts, we added polystyrene tracer particles with a diameter of \SI{1.0}{\micro\meter} to perform PIV analysis. The volume of the actomyosin cytoskeleton was $V_{act}$, the volume of the buffer solution (A50 buffer) was $V_{buf}$, and the volume fraction of the diluted actomyosin solution was defined as $\phi=V_{act}/(V_{act}+V_{buf})$. Artificial cells were prepared using cytoplasmic extracts diluted at $\phi$ = 0\%, 20\%, 30\%, 40\%, 50\%, 60\%, and 80\% in solution. Figure 2 presents representative cases of $\phi$ = 0\%, 40\%, 60\% and 80\% (Supplementary Movie 2). In the artificial cells with high dilution rates $\phi$ = 0\% and 40\%, the particles were uniformly distributed, whereas in the artificial cells with $\phi$ = 60\% and 80\%, the particles were transported to the center, and cluster formation was observed (Fig. \ref{fig2}(a)). PIV analysis revealed that at $\phi$ = 0\%, no clear velocity field appeared, with a disordered velocity and direction (Fig. \ref{fig2}(b)), reflecting Brownian motion. In contrast, in an artificial cell with $\phi$ = 60\%, the actin flow near the cluster was slightly biased toward the center of the artificial cells. The actin flow at $\phi$ = 80\% was directed toward the center, with the highest velocity near the cluster (Fig. \ref{fig2}(b)). 

\begin{figure*}[tb]
\begin{center}
\includegraphics[width=14cm]{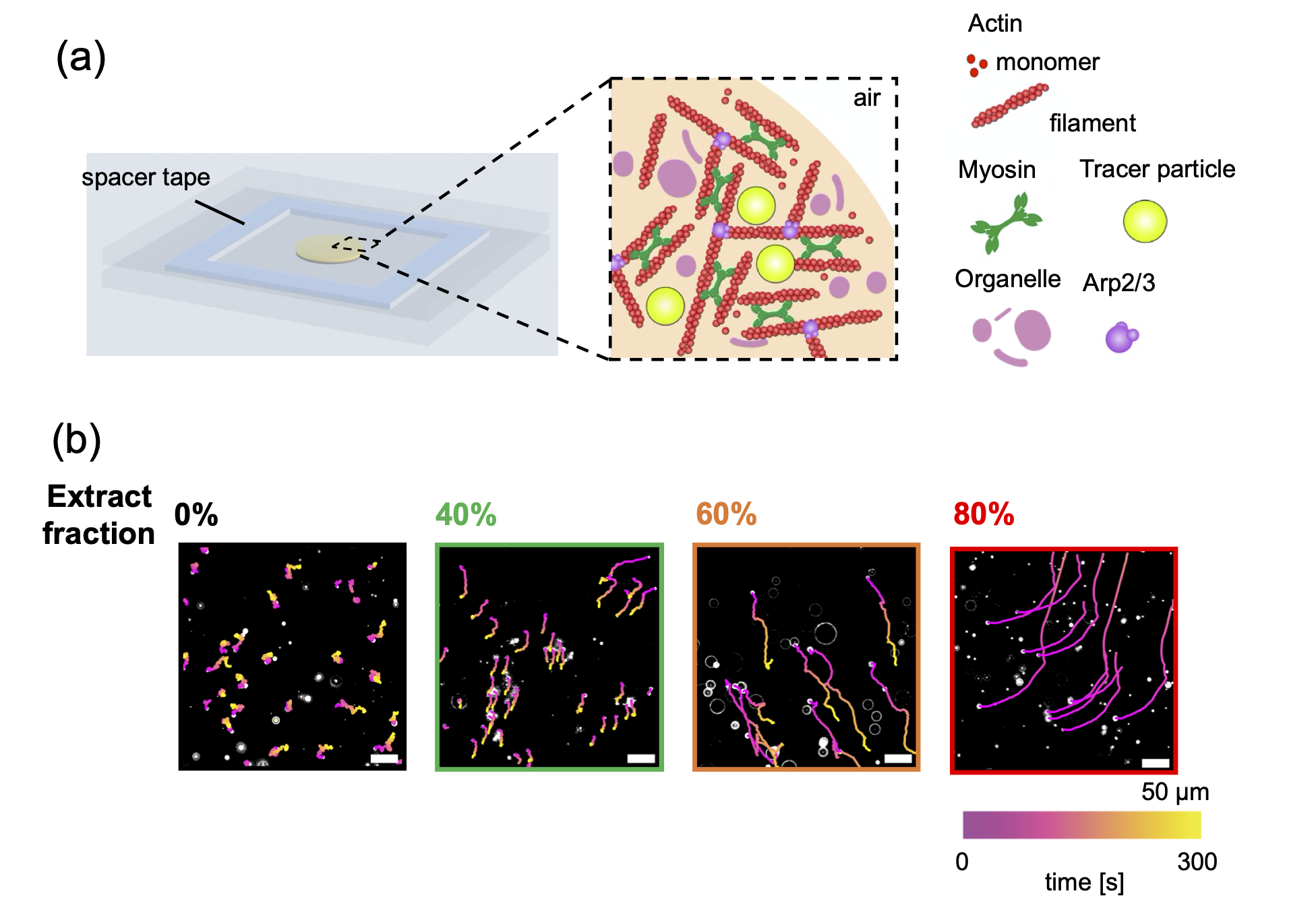}
\caption{{\footnotesize \textbf{Actin flow in bulk actomyosin solution.} (a) Schematic of an experiment to visualize bulk actin flow using fluorescent tracer particles without compartmentalization in artificial cells. (b) Microscopic images of fluorescent tracer particles across various actomyosin cytoskeleton fractions ($\phi=0$\%, 40\%, 60\%, and 80\%). Particle trajectories were analyzed using particle tracking (refer to Methods section). The color in each trajectory represents the elapsed time during tracking. Scale bar: \SI{50}{\micro\meter}.}}\label{fig3}
\end{center}
\end{figure*}

We further examined the spatial profile of actin flow by decomposing the velocity into radial ($v_r$) and tangential ($v_{\theta}$) components at the same radial distances from the center and calculated the average of each component. The radial velocity of the diluted actomyosin solution ($\phi$ = 0\%, 20\%, 30\%, 40\%) was approximately \SI{0.02}{\micro\meter\per\second}, which is comparable to the velocity of Brownian motion (Fig. \ref{fig2}(c)). At $\phi$ = 60\%, $v_r$ increased slightly near the center, reaching a maximum at about $r=$\SI{25}{\micro\meter} from the center (Fig. \ref{fig2}(c)). At a higher fraction $\phi$ = 80\%, $v_r$ exhibited a peak velocity within approximately $r=$33-\SI{50} {\micro\meter} from the center, which was \SI{0.08}{\micro\meter\per\second} (Fig. \ref{fig2}(c)). Conversely, $v_{\theta}$ remained below \SI{0.02}{\micro\meter\per\second} across all fractions $\phi$ = 0\%, 20\%, 30\%, 40\%, 50\%, 60\%, and 80\%, indicating an absence of rotational flow (Fig. \ref{fig2}(d)).

Actin flow toward the center drives cluster formation, where organelles and the actomyosin cytoskeleton accumulate. Consequently, we explored whether a positive correlation exists between the amount of actomyosin in the artificial cell and the cluster size when a sufficient fraction of actomyosin is present. Accordingly, we measured the diameters of the artificial cells $D_{drop}$ (150 - \SI{300}{\micro\meter}) and the clusters $D_{clust}$. Stable clusters formed in droplets at $\phi$ = 60\%, and a visible cluster formed at $\phi$ = 80\%, which showed a proportional increase with the size of the artificial cells (Fig. \ref{fig2}(e)). This result aligns with the mechanism whereby the actomyosin cytoskeleton and cytoplasmic contents, driven by actin flow, accumulate and enlarge the formed cluster, correlating with the increase in the size of the artificial cells.

We then investigated whether the actomyosin cytoskeleton in solution could generate flow solely through spontaneous force generation. Accordingly, we investigated the density-dependent dynamics of actin flow using a bulk solution system where the diluted cytoplasmic extract was placed directly in the chamber without confinement. We prepared a bulk of the diluted actomyosin solution by dropping \SI{6}{\micro\liter} of the solution onto a glass slide coated with bovine serum albumin (BSA), which prevents non-specific adsorption of myosin and actin proteins onto glass, maintaining a height of \SI{250}{\micro\meter} with a spacer tape. We referred to this as the "bulk" actomyosin solution (Fig. \ref{fig3}(a)). Actin flow in the bulk solution was visualized by tracking fluorescent tracer particles. We recorded the motion of tracer particles at \SI{1.0}{\s} intervals, and trajectories were obtained from the centers of the particles in bulk solutions of different actomyosin fractions (Fig. \ref{fig3}(b) and Supplementary Movie 3). At $\phi$ = 0\%, the tracer particles moved with fluctuations characteristic of thermal Brownian motion. At higher fractions $\phi$ =40\%, 60\%, and 80\%, the trajectories of the tracer particles became ballistic, moving with slight changes in direction, indicating that increasing the actomyosin fraction induced a transition from diffusive to ballistic motion, driven by density-dependent actin flow.

\begin{figure*}[tb]
\begin{center}
\includegraphics[width=14cm]{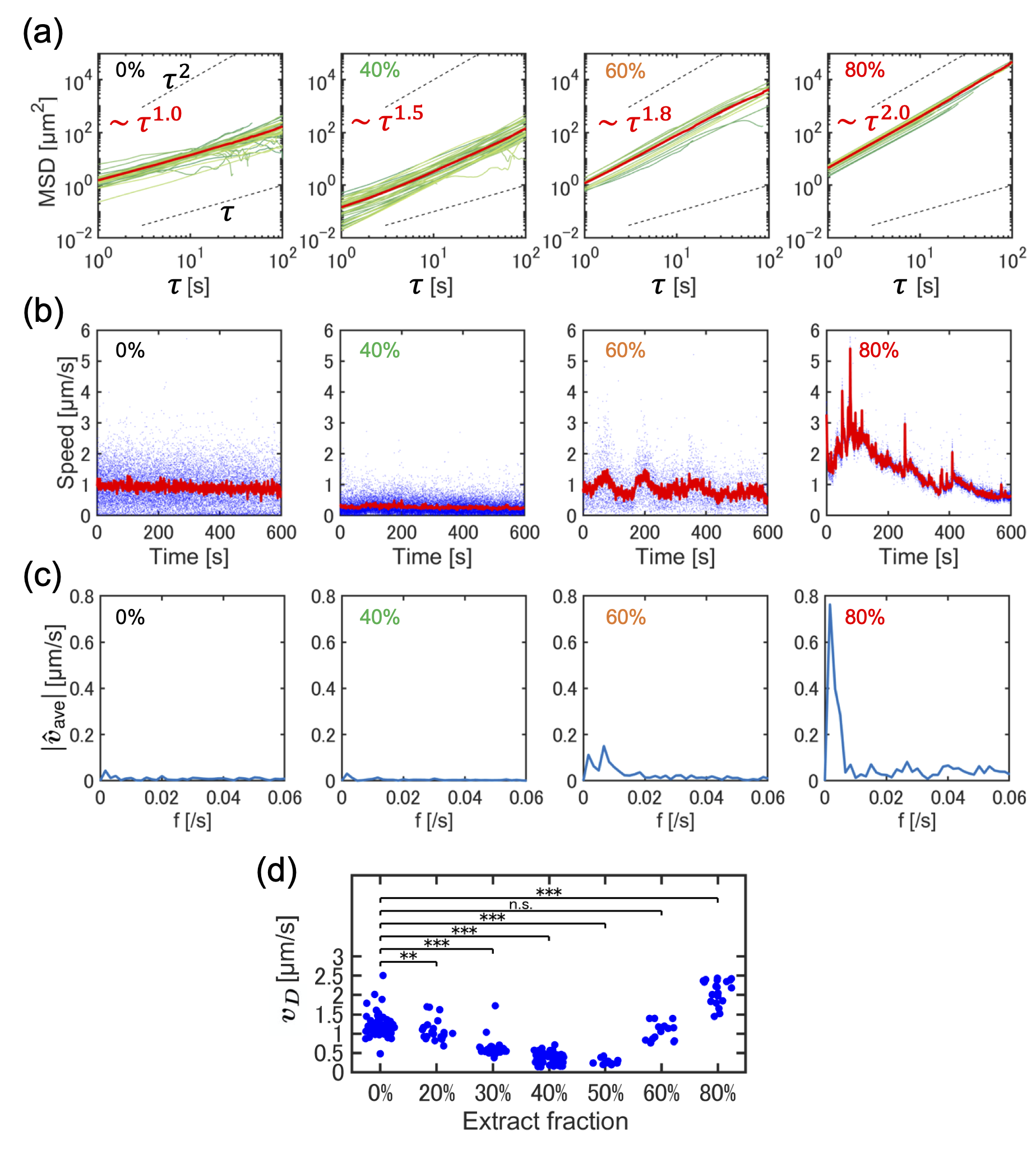}
\caption{{\footnotesize \textbf{Quantitative analysis of tracer particle trajectories in actomyosin cytoskeletons in bulk conditions.} (a) Mean squared displacement of tracer particles tracked at different actomyosin fractions ($\phi=0$\%, 40\%, 60\%, and 80\%). The green line represents single particle-tracking data, and the red line represents the averaged value. (b) Time evolution of the instantaneous speed of tracer particles. The blue scatter points represent the speed of a single particle, and the red line represents the average speed over multiple particles simultaneously. (c) Fourier transform of the time evolution of the speed of tracer particles, with a peak indicating an oscillatory change in speed appearing at $\phi$ = 60\% . (d) Density dependence of the diffusion speed $v_D$ of tracer particles ($\phi$ = 0\%, 20\%, 30\%, 40\%, 50\%, 60\% and 80\%). Statistical analysis was performed using the Mann–Whitney U test. *** indicates significance $p<0.001$, ** indicates $p<0.01$.}}\label{fig4}
\end{center}
\end{figure*}

The motion of the tracer particles is further analyzed by calculating the mean squared displacement (MSD) $\langle (\Delta r(\tau))^2\rangle$, where $\Delta r(\tau)=r(t+\tau)-r(t)$ is the distance traveled in the delay time $\tau$. We fitted the MSD curve using $\langle (\Delta r(\tau))^2\rangle \propto \tau^\alpha$ where $\alpha$ is a positive constant, allowing us to classify the dynamics of the tracer motion. When $\alpha=1$, indicating MSD $\propto \tau^{1.0}$, the tracer motion is simple diffusion with a randomly changing direction. At $\alpha = 2$, indicating $\tau^{2.0}$, the tracer displacement exhibits ballistic motion, whereas at $1<\alpha<2$, the tracer motions are super-diffusive with a partially correlated change in direction. Our experimental data revealed that at $\phi=0$\% without the cytoplasmic extract, which served as the control, the MSD was proportional to $\tau^{1.0}$, indicating simple diffusion (Fig. \ref{fig4}(a)). For actomyosin fractions ranging from $\phi=40$\% to $\phi=60$\%, super-diffusive motion appeared at $\alpha=1.5$ and $\alpha=1.8$, respectively. Furthermore, at $\phi=80$\% the tracer motion became ballistic with $\alpha=2.0$ (Fig. \ref{fig4}(a)).

\begin{figure*}[tb]
\begin{center}
\includegraphics[width=14cm]{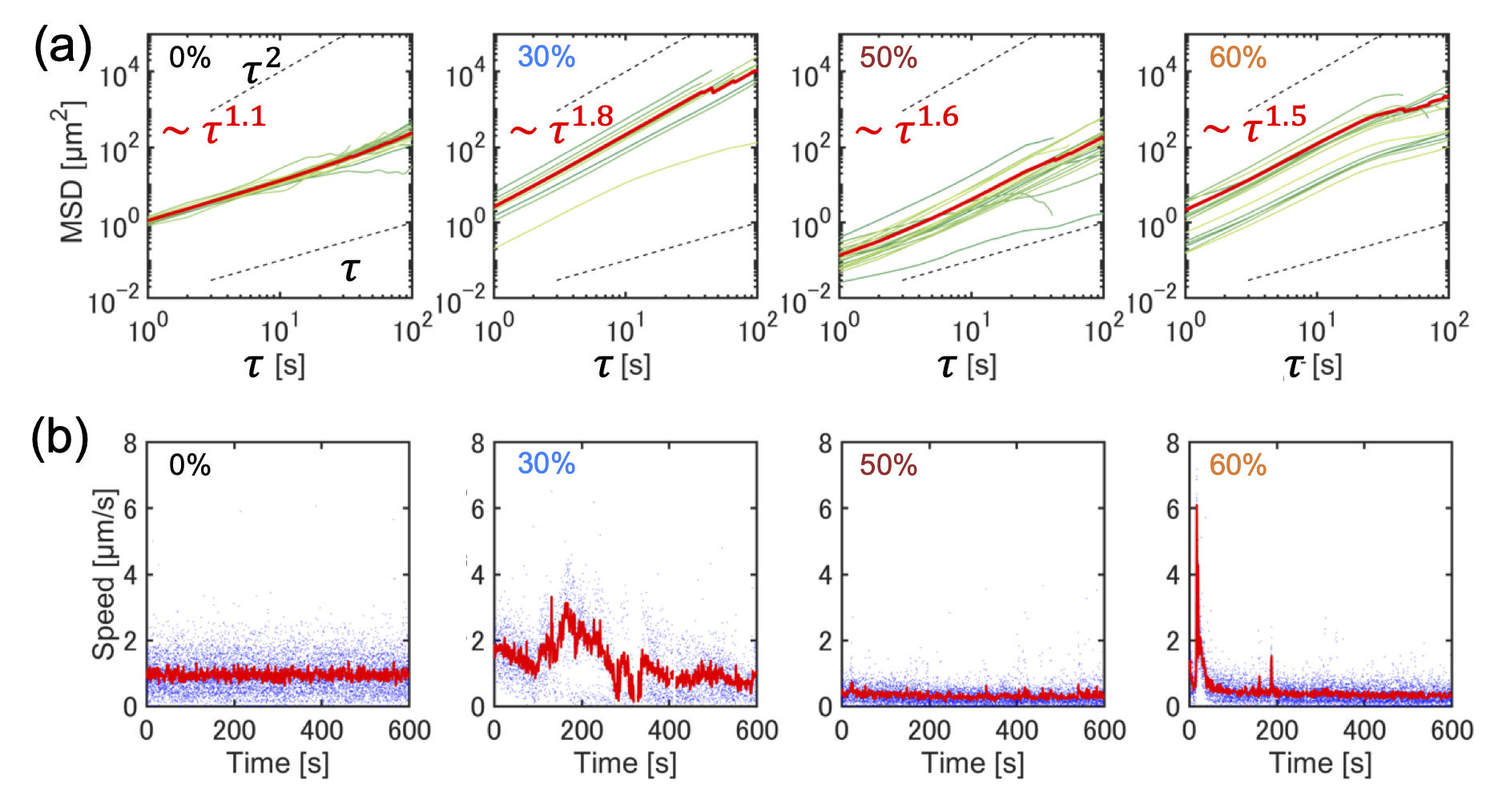}
\caption{{\footnotesize \textbf{Quantitative analysis of tracer particle trajectories in chemically-perturbed actomyosin cytoskeleton solution in bulk condition.} Actomyosin extract was chemically treated with 300 nM calyculin A (CalA-treated) to enhance myosin contractility. (a) Mean squared displacement of tracer particles tracked in various CalA-treated actomyosin fractions at representative $\phi=0$\%, 30\%, 50\%, and 60\%. We note that the MSD curves at  $\phi=60$\% are separated into two groups, actively transported particles and slowly transported ones. The mean MSD curve (red solid line) was acquired by averaging over actively transported particles. (b) Time evolution of the instantaneous speed of tracer particles in CalA-treated actomyosin solution.}}\label{fig5}
\end{center}
\end{figure*}

Next, we investigated the transition from diffusive dynamics to persistent motion of particles under actin flow by focusing on the speed of the tracer particles. We calculated the instantaneous speed $v(t)$ at time $t$ from the displacement of the particle position at intervals of 1 s (Fig. \ref{fig4}(b)). We defined the mean speed $v_{ave}(t)$ by averaging $v(t)$ over particle population and employed it for the following comparative analysis. For the buffer solution at $\phi$ = 0\% and the diluted actomyosin solution at $\phi$ = 40\%, $v_{ave}(t)$ was almost constant over time. The speed at $\phi$ = 40\% was slightly lower than that at $\phi$ = 0\%, suggesting suppressed particle displacement owing to increased viscosity with a higher fraction of the cytoplasmic extract. For the solution at $\phi$ = 60\%, $v_{ave}(t)$ exhibited periodic changes, increasing and decreasing with time (Fig. \ref{fig4}(b)), reflecting periodic stop-and-go motion. At a higher fraction, $\phi$ = 80\%, $v_{ave}(t)$ showed high initial values but then decayed with oscillatory speed fluctuations (Fig. \ref{fig4}(b)). This gradual decrease in speed is attributable to the depletion of the actomyosin cytoskeleton from the bulk solution through rapid actin flow. To confirm the periodicity of the speed, we examined the Fourier transform of $v_{ave}(t) - \langle v_{ave} \rangle$ where $\langle v_{ave} \rangle$ is the mean speed averaged over time, defined as $\hat{v}_{ave}(f)$, and observed that the particle speed at $\phi$ = 60\% exhibited regular changes in frequency $f$ = \SI{0.063}{\per\second} and period $T$ = $1/f$ = \SI{150}{\second} (Fig. \ref{fig4}(c)). Additionally, the speed fluctuation at $\phi$ = 80\% showed several peaks in the frequency space that fluctuated over periods shorter than \SI{150}{\second}.

Analyzing the dynamics of tracer particles among various fractions $\phi$ is crucial; therefore, we defined the diffusion speed $v_D = \sqrt{\langle (\Delta r(\tau))^2\rangle}/\tau$ at delay time $\tau$ = \SI{1.0}{\s}. For the solutions at $\phi$ = 20\%, 30\%, 40\% and 50\%, $v_D$ was smaller than that of normal diffusion at $\phi$ = 0\%, suggesting that even diluted fractions of actomyosin extract gradually increase the viscous friction against Brownian motion. For the solution at $\phi$ = 60\%, $v_D$ began to increase again and showed a substantial increase at $\phi$ = 80\% to \SI{1.5}{\micro\meter\per\second} (Fig. \ref{fig4}(d)). This suggests that as the fraction of cytoplasmic extract increased, Brownian motion was suppressed due to increased viscosity from higher overall protein concentrations. However, as the fraction of actomyosin extract further increased, tracer particles driven by actin flow moved faster than that via Brownian motion.

To further elucidate the effects of actomyosin contractility, we conducted a molecular perturbation experiment by adding calyculin A (Cal A), an inhibitor of myosin II phosphorylation that enhances the contractile force \cite{henson2003}. We repeated the particle-tracking experiment in a bulk system of diluted cytoplasmic extract solution containing 300 nM Cal A. The MSD of tracer particles showed that $\phi$ = 0\% exhibited behavior similar to that of simple diffusion, whereas even at a relatively low fraction $\phi$ = 30\%, it approached ballistic behavior, represented by $\tau^{1.8}$ (Fig. \ref{fig5}(a)). This indicates that the CalA-induced upregulation of contractile force can drive near-ballistic motion even in dilutions of cytoplasmic extracts at relatively low fractions. The speed of the tracer particles $v(t)$ at $\phi$ = 30\% increased by \SI{2.0}{\micro\meter\per\second} with fluctuations (Fig. \ref{fig5}(b)).

For $\phi$ = 50\% and 60\%, the slope of the MSD was smaller than that for $\phi$ = 30\% (Fig. \ref{fig5}(a)), indicating that the ballistic motion associated with actin flow was suppressed. Consistent with this result, $v(t)$ at $\phi$ = 50\% and 60\% was reduced by \SI{0.6}{\micro\meter\per\second}, comparable with that induced by thermal fluctuations (Fig. \ref{fig5}(b)). These chemical perturbation analyses suggest that actomyosin contractile force can drive ballistic tracer diffusion accompanied by actin flow; however, its strength should be maintained within a physiological range to support continuous flow.

\section{Discussion}
In this study, we experimentally analyzed the occurrence of a density-dependent flow of actin. The contraction of cytoskeletal fibers is driven by the force generation of myosin molecular motor above the critical density of $\phi=60$\%, accompanied by actin flow. This occurred at comparable levels of the extract fraction in both the confined systems of artificial cells and in the bulk solution. This flux oscillates near the transition point but is suppressed at higher myosin contractility levels, suggesting that instability due to actomyosin cytoskeleton contraction is involved. In artificial cells, periodic actin waves originate from the boundary where nucleation factors promote actin network formation, resulting in increased actin concentration at the boundary and contraction toward the center of the artificial cells. Such periodic actin waves can be observed when the actomyosin concentration is sufficiently high \cite{sakamoto2023state, krishna2024}. However, the oscillatory actin flow found in this study was at a dilute density near the transition from a uniform static state to advective flow. Therefore, the oscillatory flow mechanism reflects inherent mechanical properties. Furthermore, the period of the oscillatory actin waves was approximately \SI{60}{\second} \cite{sakamoto2020tug,sakamoto2023state,krishna2024}, but the period of the speed oscillation was longer, \SI{150}{\second}, at $\phi$ = 60\% in the bulk solution. Oscillatory actin flow was observed at a lower density than that in previous studies, and the actin network takes longer to grow, which is consistent with the involvement of the actin polymerization rate in the onset of actin flow.

As actomyosin density increased, the speed of actin flow (diffusion speed of tracer particles) initially decreased. Upon reaching the critical density for flow initiation, actin flow occurred with sustained directionality. The reduction in speed at low densities may result from the actin fiber network impeding tracer particle diffusion. In contrast, the contractile force of actomyosin is insufficiently developed to overcome thermal fluctuations. This suggests that network formation of the actin filaments, associated with nucleation factors, precedes myosin contraction-driven actin flow. Given the challenge of analyzing the concentration dependence of the intracellular cytoskeleton in living cells, a reconstituted system proves advantageous. Further investigation is warranted to address the development of an active gel model incorporating the density-dependent viscoelasticity of the actin network \cite{banerjee2011,dierkes2014a} and the contractile force of myosin to explain the experimental results.

The density-dependent flow of actin provides insights into biological self-organization, where cytoskeletal deformation owing to contractile force generation occurs beyond a certain concentration range. Intracellular actomyosin is not uniformly distributed but forms cortices on the plasma membrane surface, which is covered by a thin layer of the actomyosin cytoskeleton. Such localization patterns would be relevant for coexistence in the partitioning of intracellular space, providing sufficient contractility near the membrane to deform it, whereas the actin cytoskeleton restricts flow in the low-density spaces within the cytosolic space. Thus, the density-dependent transition of actin flow may regulate the viscoelastic mechanics underlying adhesion-independent motility in membrane blebbing.

\section{Materials and Methods}
\subsection{Sample Preparation}

The frozen actomyosin extract was incubated on ice for $\SI{1.0}{\hour}$ prior to experimental use. TMR-LifeAct (final concentration \SI{1.0}{\micro\mol\per\liter}) and nocodazole (final concentration \SI{200}{\micro\mol\per\liter}) were added to the thawed extract solution and placed on ice. The extract was diluted with A50 buffer, and tracer particles (L4655 Latex beads, carboxylate-modified polystyrene, fluorescent yellow-green, solids: 0.6$\times 10^{-2}$\%) were added for flow visualization. To prepare artificial cells, \SI{1.0}{\micro\liter} of this extract was injected into \SI{10}{\micro\liter} of mineral oil (M5904, Sigma-Aldrich), and the mixed solution was agitated via finger tapping. The oil phase contained \SI{1.0}{\milli\mol\per\liter} phosphatidylcholine (eggPC; Nacalai Tesque), and mechanical agitation helped form an emulsion containing extract-in-oil droplets.

To construct a sample chamber for artificial cells, glass slides were coated with a silicone elastomer (polydimethylsiloxane, PDMS) (Sylgard 184; Dow Corning). The uncured PDMS was spread using a spin coater (MA-100, MIKASA). The PDMS-coated glass slides were cured at \SI{75}{\celsius} for \SI{1}{\hour} and then cut into three pieces on a PDMS-coated glass plate. Subsequently, \SI{3.0}{\micro\liter} of the extract-in-oil emulsion was placed in the PDMS-coated glass chamber. The height of the chamber was set to \SI{90}{\micro\meter} by inserting a spacer tape (Nichiban).

We also constructed a chamber using glass slides (Matsunami, S1111 and S1127) to prepare a bulk actomyosin solution. Approximately \SI{6.0}{\micro\liter} of the extract was placed on a BSA-coated glass slide. The height of the chamber (i.e. spacer thickness) was \SI{250}{\micro\meter}. Immediately after \SI{1.0}{\minute} of plasma cleaning, \SI{20}{\micro\liter} of phosphate buffer solution containing 0.1 mg/mL BSA (A2153, Sigma-Aldrich) was applied and spread near the center of the glass. After \SI{10}{\minute} of incubation at room temperature, the BSA solution was removed, and the chamber was washed twice with phosphate buffer.

\subsection{Microscopic Observation}
Time-lapse microscopy images of the artificial cells were captured every \SI{5.0}{\second} using an inverted microscope (IX73; Olympus) equipped with a 20× objective lens and an EM-CCD camera (iXon, Andor Technology). Time-lapse images of the bulk system were acquired every \SI{1.0}{\second} using an epifluorescence microscope (IX73, Olympus) equipped with a cooled CMOS camera (Neo-5.5-CL3; Andor Technology). The temperature of the microscope stage was maintained at $20 \pm$\SI{2}{\celsius} using a custom-made copper chamber.

\subsection{Image Processing}

Image processing analysis was performed using custom code in MATLAB. To visualize actin flow, PIV analysis was performed using OpenPIV \cite{bengida2020}. Additionally, to analyze the dynamics of tracer particles in the bulk actomyosin solution, TrackMate, implemented as a plugin for Fiji/ImageJ, was used to track the fluorescent particles mixed into the actomyosin solution \cite{sage2005,ershov2022}.

\section*{Data availability}

The authors confirm that the data supporting the findings of this study are available within the article.

\section*{Conflicts of interest}

There are no conflicts of interest to declare.

\section*{Acknowledgements}

We thank M. Miyazaki for kindly providing frozen actomyosin extract. This work was supported by Grant-in-Aid for Scientific Research(B)(23H01144), Grant-in-Aid for Challenging (Exploratory)(24K21534), Grant-in-Aid for Transformative Research Areas(A)(23H04711, 23H04599), Grant-in-Aid for Early-Career Scientists (22K14014), JST FOREST Grant JPMJFR2239, JSPS Core-to-Core Program “Advanced core-to-core network for the physics of self-organizing active matter" (JPJSCCA20230002), Sumitomo Foundation, and Joint Research of ExCELLS (23EXC205 and 24EXC206). T.K. acknowledges support from the JSPS fellowship DC1 (24KJ1796).

\bibliography{actomyosin}

\end{document}